\documentclass{ws-procs961x669}            
\begin{document}
\title{Gravitational Wave Emission by Accretion of Matter and Magnetic Deformation in Fast Rotating White Dwarfs}

\author{M. F. Sousa$^*$ and J. C. N. de Araujo$^{**}$}

\address{Divis\~{a}o de Astrof\'{i}sica, Instituto Nacional de Pesquisas Espaciais,\\
S\~{a}o Jos\'{e} dos Campos, SP/12227-010, Brazil\\
$^*$E-mail: manoel.sousa@inpe.br\\
$^{**}$E-mail: jcarlos.dearaujo@inpe.br\\
www.inpe.br}

\author{J. G. Coelho}

\address{Departamento de F\'isica, Universidade Tecnol\'ogica Federal do Paran\'a,\\
Medianeira, PR/85884-000, Brazil\\
E-mail: jazielcoelho@utfpr.edu.br}

\begin{abstract}

We discuss some aspects of Sousa et al.[\citenum{sousa2020gravitational1,sousa2020gravitational2}] concerning two mechanisms of gravitational wave (GW) emission in fast-spinning white dwarfs (WDs): accretion of matter and magnetic deformation. In both cases, the GW emission is generated by an asymmetry around the rotation axis of the star. However, in the first case, the asymmetry is due to the amount of accreted matter in the magnetic poles, while in the second case it is due to the intense magnetic field. We have estimated the GW amplitude and luminosity for three binary systems that have a fast-spinning magnetized WD, namely, AE Aquarii, AR Scorpii and RX J0648.0-4418. In addition, we applied the magnetic deformation mechanism for SGRs/AXPs described as WD pulsars. We found that, for the first mechanism, the systems AE Aquarii and RX J0648.0-4418 can be observed by the space detectors BBO and DECIGO if they have an amount of accreted mass of $\delta m \geq 10^{-5}M_{\odot }$. For the second mechanism, the three systems studied require that the WD has a magnetic field above $\sim 10^{9}$ G to emit GWs that can be detected by BBO. Furthermore, we found that some SGRs/AXPs as WD pulsars can be detected by BBO and DECIGO, whereas SGRs/AXPs as highly magnetized neutron stars are far below the sensitivity curves of these detectors.
\end{abstract}


\bodymatter

\section{Introduction}
\label{aba:sec1}

Over the last years, the astrophysical community's interest in highly magnetized white dwarfs (HMWDs) has increased. Recent results of the Sloan Digital Sky Survey (SDSS) have confirmed these white dwarfs (WDs) with surface magnetic fields ranging from $10^6$~G up to $10^9$~G [see e.g. Refs.~\citenum{2009A&A...506.1341K,2013MNRAS.429.2934K,2015MNRAS.446.4078K}]. Besides their high magnetic fields, most of them have been shown to be massive, and responsible for the high-mass peak at $1~\textrm{M}_\odot$ of the WD mass distribution \cite{2009A&A...506.1341K,1992ApJ...394..603S,2010A&A...524A..36K}. 

Typically, WDs have rotation periods of days or even years. However, recently, a WD pulsar was discovered, called AR Scorpii. This star emits from X-ray to radio wavelengths, pulsing in brightness with a period of $1.97$ min \cite{2016Natur.537..374M}. Moreover, other sources have been proposed as candidates of WD pulsars. Specific examples are AE Aquarii with a short rotation period of $P=33.08$~s \cite{2008PASJ...60..387T} and RX J0648.0-4418 (RX J0648, hereafter) that is a massive WD with $M=1.28M_\odot$ and with a very fast spin period of $P = 13.2$~s, that belongs to the binary system HD 49798/RX J0648.0-4418\cite{2009Sci...325.1222M}. Nevertheless, it is worth mentioning that the nature of RX J0648 is unclear, meaning it is not yet clearly known whether this star is a WD or a neutron star \cite{mereghetti2016,popov2017}.

Recently, from XMM-Newton observations, the authors in Ref.~\citenum{2020ApJ...898L..40L} reported that CTCV J2056-3014 is a X-ray-faint intermediate polar harboring an extremely fast-spinning WD with a coherent pulsation of 29.6 s, thus being the fastest confirmed spin in a WD [see also Ref. \citenum{2020arXiv201012441O}]. Other fast-spinning WDs have been observed more recently. As examples we can cite: V1460 Her, which is an eclipsing cataclysmic variable, with a overluminous K5-type donor star and a WD that rotates with a period of 38.9 s \cite{Ashley2020} and ZTF J190132.9+145808.7, which is a highly magnetized and rapidly rotating white dwarf, featuring a magnetic field  with strengths between 600 MG and 900 MG on its surface, and a stellar radius that is only slightly larger than the radius of the Moon \cite{caiazzo2021highly}. This WD has a rotation period of 6.94 min which is considered short as this star is an isolated WD.

Notwithstanding, several studies of magnetized and fast-rotating WDs have been done. In particular, we can highlight one involving WD pulsars in an alternative description for Soft Gamma Repeaters (SGRs) and Anomalous X-Ray Pulsars (AXPs) [see e.g. Refs.~\citenum{Malheiro/2012,Coelho/2014,2016IJMPD..2541025L,2016JCAP...05..007M}]. From this perspective, the process of energy emission released by dipole radiation in a WD can be explained by a canonical spin-powered pulsar model, since they share quite similar aspects \cite{Usov/1988,Coelho/2014}.

On the other hand, LIGO and Virgo detectors have recently made direct observations of gravitational waves (GWs) \cite{abbott2019gwtc,abbott2021gwtc}. All these GW detections are within a frequency band ranging from $10$ Hz to $1000$ Hz, which is the operating band of LIGO and Virgo. Nevertheless, as is well known, there are proposed missions for lower frequencies, such as LISA \cite{Amaro/2017,cornish2018}, whose frequency band is of $(10^{-4}-0.01)$ Hz, BBO \cite{harry/2006,yagi2011} and DECIGO \cite{kawamura/2006,yagi2017} in the frequency band ranging from $0.01$ Hz to $10$ Hz.

The generation of continuous GWs in different possibilities has already been proposed \cite{1996A&A...312..675B,2016JCAP...07..023D,2016ApJ...831...35D,2017EPJC...77..350D,Schramm/2017}. More recently, in Refs. \citenum{2019MNRAS.tmp.2346K,sousa2020gravitational1, sousa2020gravitational2}, it has been suggested that rotating magnetized WDs can emit continuous GWs with amplitudes possibly detected by upcoming GW detectors such as LISA, DECIGO and BBO. Here we revisit our two works [Refs. \citenum{sousa2020gravitational1,sousa2020gravitational2}], where we investigate two mechanisms of gravitational radiation emission in fast-rotating magnetized WD: matter accretion and magnetic deformation. In both cases, the emission in GW is produced by the asymmetry around the star's rotation axis.

\section{Gravitational emission mechanisms  }
\label{aba:sec2}

WDs might generate gravitational radiation whether they are not perfectly symmetric around their rotation axis. The huge dipole magnetic field that can make the star become oblate \cite{chandrasekhar1953s} and accretion of matter are two examples where this asymmetry can occur.

\subsection{Accretion of matter}

The GW emission is shown here for the case of a WD accreting matter via the magnetic poles which do not coincide with the rotation axis of the star. In this scenario, the system's secondary star transfers matter to the WD through an accretion column and accumulates an amount of mass on the magnetic poles.

Thus, we consider a rigid object rotating about a non-major axis ($x_{1}$, $x_{2}$, $x_{3}$) and which has a deformity about one of the major axes ($x_{1}$, $x_{2}$, $x_{3}$), where are positioned the main moments of inertia $I_{1}$, $I_{2}$ e $I_{3}$, respectively. 

With this configuration and doing $I_{1} = I_{2}$, the gravitational amplitude and luminosity are given respectively by \cite{shapiro/2008,maggiore/2008}

\begin{equation}
  h_{0_{ac}} = \frac{4G}{c^{4}} \frac{(I_{1}-I_{3}) \omega ^{2}}{r} \sin ^{2}\theta,
  \label{AmpAC2}
\end{equation}

\begin{equation}
  L_{GW_{ac}} = -\frac{2}{5} \frac{G}{c^{5}}(I_{1}-I_{3})^{2} \omega ^{6} \sin^{2}\theta 
  \\ (16 \sin^{2}\theta + \cos^{2}\theta ),
 \label{lumAC1}
\end{equation}

\noindent where, $\omega$ is the angular velocity, $\theta$ is the angle between the rotation and magnetic dipole axes and $r$ is the distance to the emitting source. 

Now, to determine the moments of inertia $I_{1}$ and $I_{3}$, we consider that the object has an amount of mass accumulated on the $x'_{3}$ axis. We reduce this system to a large sphere with two smaller spheres of matter on the $x'_{3}$ axis: one at each of the poles of the larger sphere. This would be equivalent to a WD accreting matter by the two magnetic poles. Therefore, it follows immediately that

\begin{eqnarray}
    I_{1} = \frac{2}{5} MR^{2} + 2\delta m ~R^{2}, \nonumber \\
    I_{3} = \frac{2}{5} MR^{2} + 2 \frac{2}{5} \delta m ~a^{2},
\label{I1AC}
\end{eqnarray}

\noindent where $M$ is the mass of the star, $R$ is the radius of the star, $\delta m$ is the amount of mass accumulated on one magnetic pole and $a$ its radius.

Considering that $R \gg a$ and by substituting these last expressions into Eqs. (\ref{AmpAC2}) and (\ref{lumAC1}),  one obtains

\begin{equation}
  h_{0_{ac}} = \frac{8G}{c^{4}} \frac{\delta m ~R^{2} \omega ^{2}}{r} \sin ^{2}\theta,
    \label{AmpAC4}
 \end{equation}
 
\noindent and

\begin{equation}
  L_{GW_{ac}} = -\frac{8}{5} \frac{G}{c^{5}} \delta m^{2} R^{4} \omega ^{6} \sin^{2}\theta (16 \sin^{2}\theta + \cos^{2}\theta ).
    \label{lumAC2}
 \end{equation}
 
Thereby, we find expressions for the gravitational luminosity and the GW amplitude for the case of a WD accumulating mass, which depends on the accreted mass, the distance to the source, the radius of the star and how fast it is rotating.

\subsection{Magnetic deformation}

This section deals with the deformation of the WD induced by its own huge magnetic field. Due to the combination of magnetic field and rotation, a WD can become  triaxial, presenting therefore a triaxial moment of inertia. In order to investigate the effect arising from the magnetic stress on the equilibrium of stars, let  us  introduce  the equatorial  ellipticity,  defined as \cite{shapiro/2008,maggiore/2008}

\begin{equation}
  \epsilon = \frac{I_{1} - I_{2}}{I_{3}}.
 \label{elipticidade}
\end{equation}

\noindent where $I_{1}$, $I_{2}$ and $I_{3}$ are main moments of inertia with respect to the ($x$, $y$, $z$) axes, respectively.

If the star rotates around the $z-$axis, then it will emit monochromatic GWs with a frequency twice the rotation frequency, $f_{rot}$, with amplitude given by \cite{shapiro/2008,maggiore/2008}

\begin{equation}
   h_{0_{df}} = \frac{16 \pi^{2} G}{c^{4}}  \frac{I_{3} f_{rot}^{2}}{r} \, \epsilon,
\label{Ampdef2}
\end{equation}

\noindent and luminosity as follows: 

 \begin{equation}
   L_{GW_{df}} = - \frac{32}{5} \frac{G}{c^{5}} I_{3}^{2} \epsilon^{2} \omega_{rot}^{6}.
\label{Lumidef1}
\end{equation}

On the other hand, recall that the ellipticity of magnetic origin can be written as \cite{2000A&A...356..234K,2006A&A...447....1R}

\begin{equation}
\epsilon = \kappa \frac{B_{s}^{2} R^{4}}{G M^{2}},
    \label{excentridade}
\end{equation}

\noindent where, $B_{s}$ is the magnetic field strength on the star’s surface and $\kappa$ is the distortion parameter, which depends on the magnetic field configuration and equation of state (EoS) of the star.

Now, substituting this last equation into Eqs.~(\ref{Ampdef2}) and (\ref{Lumidef1}) and considering $I_3 = 2MR^{2}/5 $, one immediately obtains that

\begin{equation}
   h_{df} = \frac{32 \pi^{2} }{5 c^{4}}  \frac{R^{6} f_{rot}^{2}}{r M} \kappa B_{s}^{2},
\label{Ampdef3}
\end{equation}

\begin{equation}
   L_{GW_{df}} = - \frac{2^{13} \pi^{6} }{5^3 c^{5}} \frac{ R^{12} f_{rot}^{6}}{G M^{2}}\kappa^2 B_{s}^{4}.
\label{Lumidef2}
\end{equation}

Note that the two equations just above depend on the rotation frequency and the magnetic field strength.

In contrast, the GW amplitude can also be written as a function of the variation of the star's rotation frequency $\dot{f}_{rot}$. In this case, we must consider that a part of the spindown luminosity is converted into GWs. Thus, we can infer an efficiency, $\eta_{df}$, for the variation of the rotation frequency as $\dot{\bar{f}}_{rot} = \eta_{df} \dot{f}_{rot}$, such that $\dot{\bar{f}}_{rot}$ can be interpreted as the part of $\dot{f}_{rot}$ related to the GW brake. Hence, the GW amplitude can be written as follows

\begin{equation}
  h_{0}^{sd} = \left ( \eta_{df} \, \frac{5}{2} \frac{G}{c^{3}}  \frac{I_{3} \dot{f}_{rot}}{r^{2} f_{rot}} \right )^{1/2}.
    \label{Ampsp2}
 \end{equation}

\section{Gravitational waves from rapid rotation white dwarfs}
\label{aba:sec3}

\subsection{Accretion of matter}

Considering the scenario of an amount of mass accumulated on the magnetic poles, we apply Eq.~(\ref{AmpAC4}) for the three binary systems that have fast-spinning WDs: AE Aqr, AR Sco and RX J0648. The parameters for the systems are listed in Table \ref{tabla: sisbin}.

From Eq.~(\ref{AmpAC4}), one notes that the amplitude depends on the amount of mass accumulated; however, it is not easy to predict how much matter may have been accreted to WD and how much has been dispersed on its surface. Thus, for this analysis, we assign four values for the mountain of matter for the analyzed systems: $\delta m =$ ($10^{-3} M_{\odot } $, $10^{-4}M_{\odot } $, $10^{-5}M_{\odot } $, $10^{-6}M_{\odot } $) [see Refs. \citenum{Welsh/1998,warner2003,Hellier2001} for details about accretion in WDs]. Besides that, we consider that the angle between the magnetic and rotation axes is $\theta = 30^{\circ}$.

\begin{table}
\tbl{Parameters of 3 binary systems: Period ($P$), spindown ($\dot{P}$), WD radius (R) and distance to Earth ($r$).}
{\begin{tabular}{lccccc}
\hline
 \textbf{Systems} & \begin{tabular}[c]{@{}c@{}}$P$\\ (s)\end{tabular} & \begin{tabular}[c]{@{}c@{}}$\dot{P}$\\ ($10^{-15}$ s/s)\end{tabular} &  \begin{tabular}[c]{@{}c@{}}$R$\\ ($10^{8}$ cm)\end{tabular} & \begin{tabular}[c]{@{}c@{}}$r $\\ (pc)\end{tabular} \\ \hline
 AE Aqr$^{a}$  & 33.08 & 56.4 & 7.0 & 100 \\ 
 AR Sco$^{b}$ & 118.2 & 392 & 7.1 & 116 \\
 RX J0648$^{c}$ & 13.18 & 6.0 & 3.0 & 650 \\ \hline
\end{tabular}
}
\begin{tabnote}
$^{\text a}$ see~\citenum{CHOIYI/2000};
$^{\text b}$ see~\citenum{Schramm/2017};
$^{\text c}$ see~\citenum{mereghetti2011}.\\
\end{tabnote}
\label{tabla: sisbin}
\end{table}

\begin{figure}
\begin{center}
\includegraphics[width=8cm, height=5.3cm]{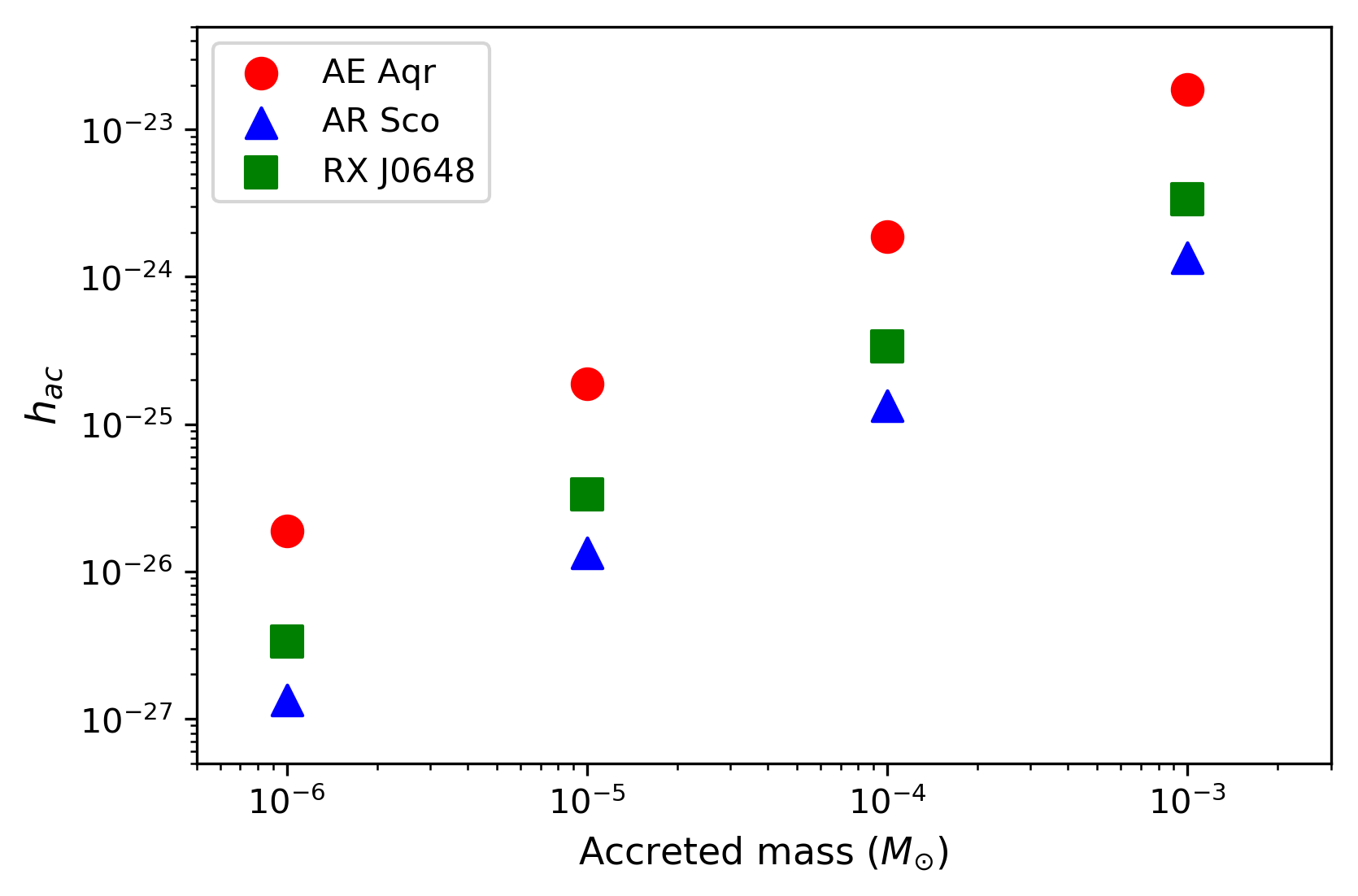}
\caption{GW amplitude as a function of accreted mass to AE Aqr, AR Sco and RX J0648}
\label{fig:1} 
\end{center}
\end{figure}

Assuming these values for $\delta m$ and the parameters listed in Table \ref{tabla: sisbin}, we obtain $h_{0_{ac}}$ for the three systems, which are shown in Figure \ref{fig:1}.

\begin{figure}
\begin{center}
\includegraphics[width=8cm, height=5.3cm]{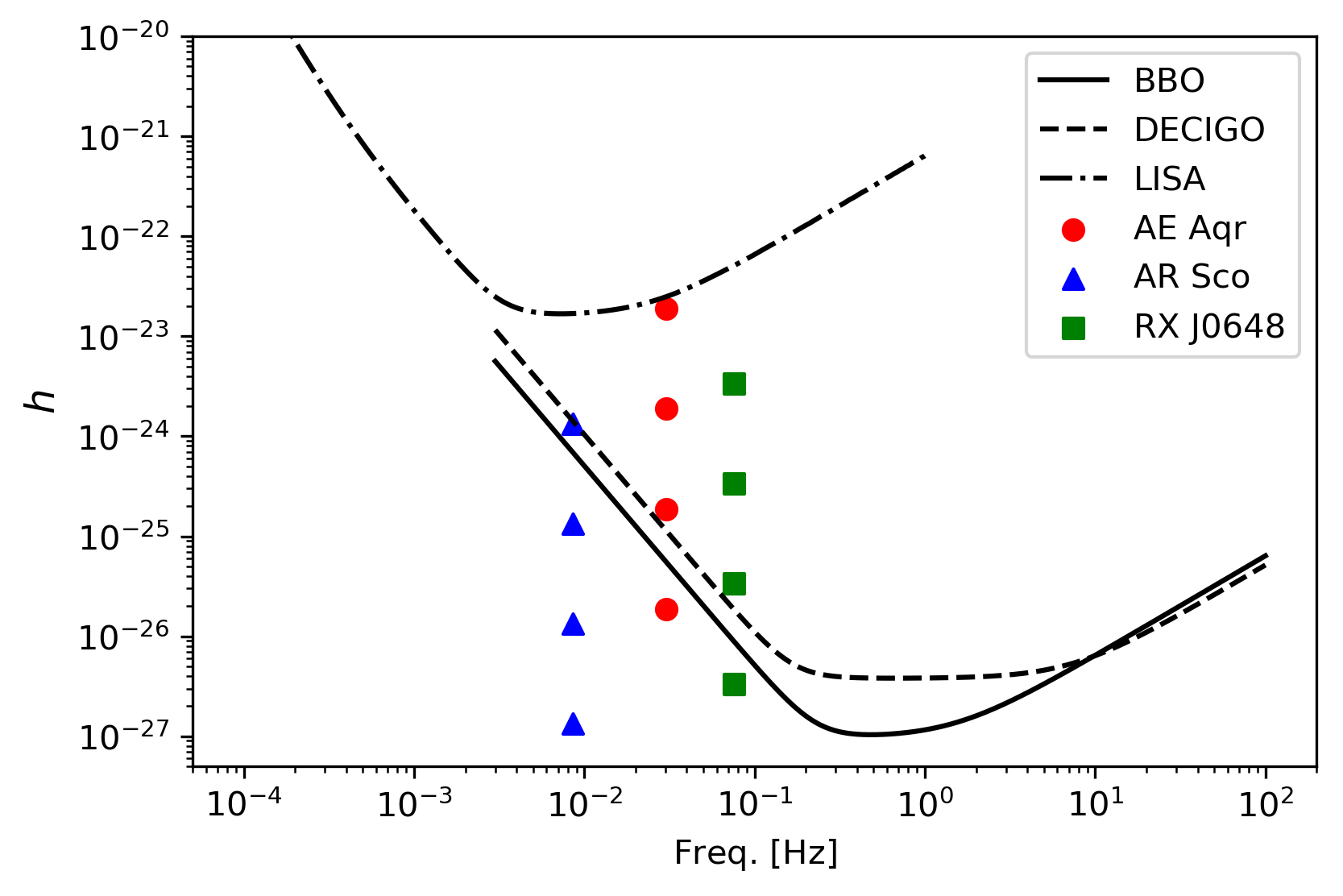}
\caption{GW amplitude for AE Aqr, AR Sco and RX J0648 for different values of mass ($10^{-3}  M_{\odot}$, $10^{-4}  M_{\odot}$, $10^{-5}  M_{\odot}$, $10^{-6}  M_{\odot}$, from top to bottom) and the sensitivity curves for LISA, BBO and DECIGO for $SNR = 8$ and integration time of $T = 1$ yr.}
\label{fig:2} 
\end{center}
\end{figure}

With the amplitude estimations, we compare them with the sensitivity curves of the gravitational wave space detectors. This outcome is shown in Figure \ref{fig:2} where we have the sensitivity curve for LISA, BBO and DECIGO with a signal-to-noise ratio (SNR) of  8 and an integration time of $T = 1$ yr. We notice from this figure that AE Aqr and RX J0648 are good candidates to be detected by BBO and DECIGO if they have an accumulated mass of  $\delta m \geq 10^{-5}M_{\odot }$. For the AR Sco system, the gravitational radiation emitted by this process would hardly be able to be detected by the three space instruments. This system would need to have a very high mass mountain of $ \sim 10^{-3}M_{\odot }$ to be above, for example, the sensitivity curve of the BBO detector.

Now, we consider the efficiency of this mechanism with respect to the rotational energy rate lost by the systems. To do this, we consider the efficiency of the process ($\eta _{acr} = L_{GW_{acr}}/L_{sd}$) for the four $\delta m$'s considered above, i.e., how much of the spindown luminosity is converted to gravitational luminosity for every $\delta m$ (see Table \ref{tabla:eficie_acr}). We find that the contribution of gravitational luminosity to the spindown luminosity is irrelevant, since, for the four values of $\delta m$ adopted, the efficiency $\eta _{acr} \ll 1$. Thereby, the contribution of gravitational luminosity to the spindown luminosity is negligible when we consider this mechanism.

\begin{table}
\tbl{The efficiency of the mechanism of GWs due to the amount of mass accumulated at the WD magnetic poles for different values of $\delta m$.}
{
\begin{tabular}{cc}
\hline
\multicolumn{2}{c}{\textbf{AE Aquarii}} \\ \hline
\begin{tabular}[c]{@{}c@{}}$\delta m$\\ ($M_{\odot }$)\end{tabular} & \begin{tabular}[c]{@{}c@{}}$\eta _{acr}$\\ ($L_{GW_{acr}}/L_{sd}$)\end{tabular} \\ \hline
$10^{-3}$ & $1.02 \times 10^{-2}$ \\ 
$10^{-4}$ & $1.02 \times 10^{-4}$ \\ 
$10^{-5}$ & $1.02 \times 10^{-6}$ \\ 
$10^{-6}$ & $1.02 \times 10^{-8}$ \\ \hline
\end{tabular}

\begin{tabular}{cc}
\hline
\multicolumn{2}{c}{\textbf{AR Scorpii}} \\ \hline
\begin{tabular}[c]{@{}c@{}}$\delta m$\\ ($M_{\odot }$)\end{tabular} & \begin{tabular}[c]{@{}c@{}}$\eta _{acr}$\\ ($L_{GW_{acr}}/L_{sd}$)\end{tabular} \\ \hline
$10^{-3}$ & $3.41 \times 10^{-5}$ \\ 
$10^{-4}$ & $3.41 \times 10^{-7}$ \\ 
$10^{-5}$ & $3.41 \times 10^{-9}$ \\ 
$10^{-6}$ & $3.41 \times 10^{-11}$ \\ \hline
\end{tabular}

\begin{tabular}{cc}
\hline
\multicolumn{2}{c}{\textbf{RX J0648}} \\ \hline
\begin{tabular}[c]{@{}c@{}}$\delta m$\\ ($M_{\odot }$)\end{tabular} & \begin{tabular}[c]{@{}c@{}}$\eta _{acr}$\\ ($L_{GW_{acr}}/L_{sd}$)\end{tabular} \\ \hline
$10^{-3}$ & $0.175$ \\ 
$10^{-4}$ & $1.75 \times 10^{-3}$ \\ 
$10^{-5}$ & $1.75 \times 10^{-5}$ \\ 
$10^{-6}$ & $1.75 \times 10^{-7}$ \\ \hline
\end{tabular}

}
\label{tabla:eficie_acr}
\end{table}

\subsection{Magnetic deformation}

\begin{figure}
\begin{center}
\includegraphics[width=8cm, height=5.3cm]{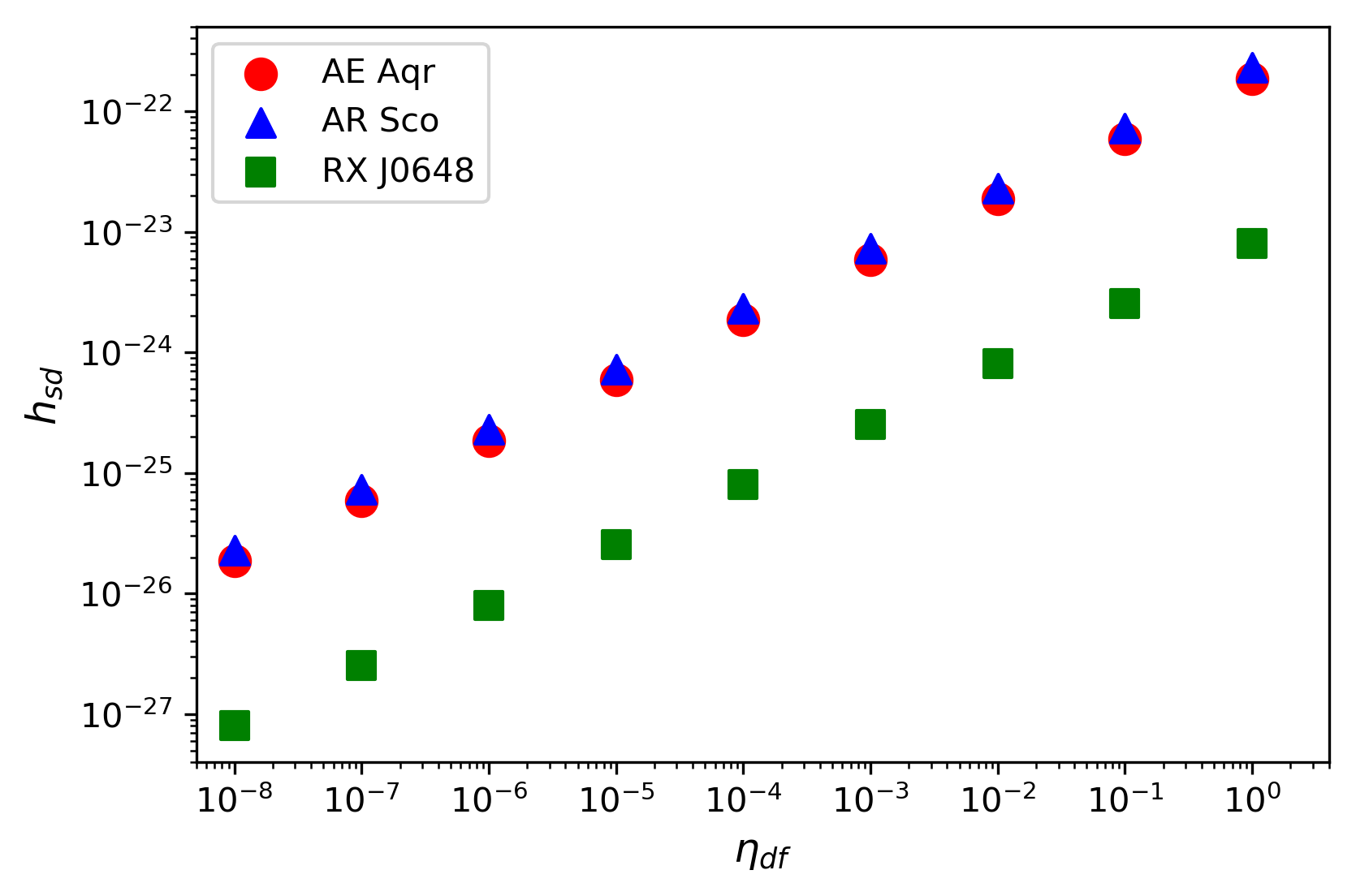}
\caption{GW amplitude for different values of efficiency ($\eta_{df} = L_{GW_{def}}/L_{sd}$) to AE Aqr, AR Sco and RX J0648}
\label{fig:3} 
\end{center}
\end{figure}

Here, we consider the generation of GWs due to the deformation of the WD structure of the same binary systems (AE Aqr, AR Sco and RX J0648) caused by their own intense magnetic field. For this, we use Eq. (\ref{Ampsp2}) to calculate the GW amplitude as a function of the efficiency $\eta_{df} = L_{GW_{def}}/L_{sd}$. The GW amplitudes are shown in Figure \ref{fig:3} as a function of $\eta_{df}$, where we use the parameters of Table \ref{tabla: sisbin} for all 3 systems.

\begin{figure}
\begin{center}
\includegraphics[width=8cm, height=5.3cm]{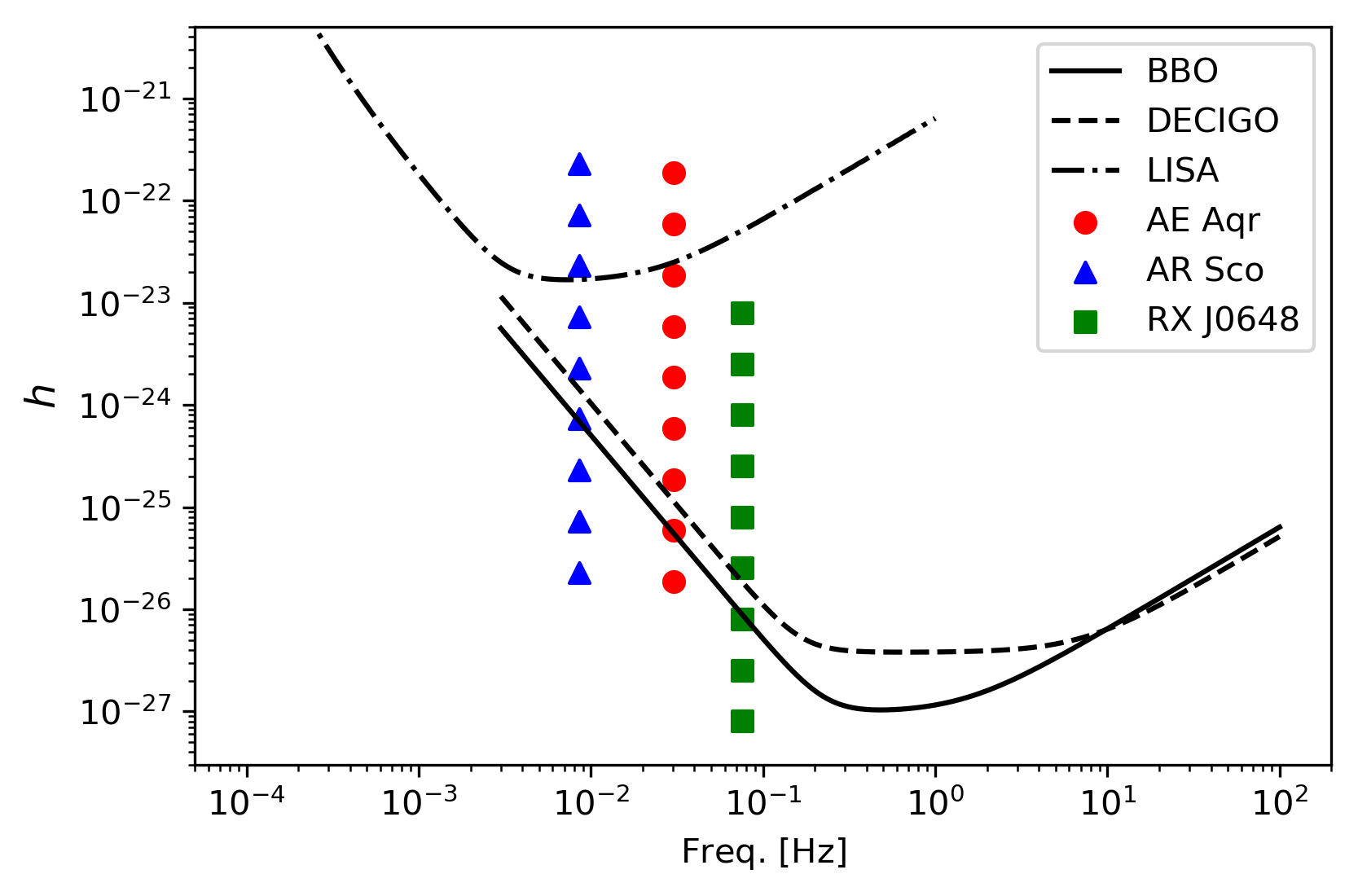}
\caption{GW amplitudes as presented in Figure \ref{fig:3} compared to the sensitivity curves of LISA, BBO and DECIGO for $SNR = 8$ and integration time of $T = 1$ year. Here, the efficiency values ($1$, $10^{-1}$, $10^{-2}$, $10^{-3}$, $10^{-4}$, $10^{-5}$, $10^{-6}$, $10^{-7}$ and $10^{-8}$) are displayed from top to bottom.}
\label{fig:4} 
\end{center}
\end{figure}

From Figure \ref{fig:4}, we plot the GW amplitudes inferred in Figure \ref{fig:3}, together with sensitivity curves for LISA, BBO and DECIGO for one year of integration time and $SNR = 8$. It is worth noting that all three systems are detectable by BBO and DECIGO as long as AE Aqr has an efficiency $\eta_{df} \geq  10^{-6}$, AR Sco an efficiency $\eta_{df} \geq  10^{-4}$ and RX J0648 an efficiency $\eta_{df} \geq  10^{-5}$. Thus, even if the GWs have a small contribution to the spindown of these systems, they can emit GWs with amplitudes that can be detected by the space antennas.

Nevertheless, it is interesting to know what the value of the magnetic field needed to produce these detectable amplitudes. Thus, we calculate the magnetic field strength so that these sources can be detected by BBO, which is the most sensitive instrument of the three considered in the present work. To do so, we use Eq. (\ref{Ampdef3}) together with the minimum efficiency for which each system is detectable by this instrument and we adopted $\kappa\simeq 10$ [see Ref. \citenum{1954ApJ...119..407F}]. Table \ref{tabla:eficie_def2} shows theses results. Notice that the systems must have WDs with high magnetic fields, around $(10^{9} - 10^{10})$ G, which are about two orders of magnitude larger than the canonical model of WD pulsars.

In addition, we can further calculate the GW amplitude by considering the upper limit values of the magnetic field strength, $B_{dip}$, inferred from the canonical model of WD pulsars. Table \ref{tabla:eficie_def3} presents the results of this study. Notice that the amplitudes of the GW shown in this Table is very small to be observed by the space detectors, since they are well below the sensitivities of these detectors. In other words, the space detectors will not be able to detect these sources when considering the upper limit of the magnetic field strength.

\begin{table}
\tbl{Minimum efficiency for the sources to be detected by the BBO detector along with the required magnetic field strength.}
{\begin{tabular}{lccc}
\hline
\multicolumn{4}{c}{\textbf{Minimum efficiency detected by BBO}} \\ \hline
\multicolumn{1}{l}{\textbf{Systems}} & \multicolumn{1}{c}{$\eta _{df}$} & \multicolumn{1}{c}{$h_{def}$} & \multicolumn{1}{c}{$B$ (G)} \\ \hline
\multicolumn{1}{l}{AE Aqr} & \multicolumn{1}{c}{$10^{-6}$} & \multicolumn{1}{c}{$1.9 \times 10^{-25}$} & \multicolumn{1}{c}{$2.8 \times 10^{9}$} \\ 
\multicolumn{1}{l}{AR Sco} & \multicolumn{1}{c}{$10^{-4}$} & \multicolumn{1}{c}{$2.3 \times 10^{-24}$} & \multicolumn{1}{c}{$3.6 \times 10^{10}$} \\ 
\multicolumn{1}{l}{RX J0648} & \multicolumn{1}{c}{$10^{-5}$} & \multicolumn{1}{c}{$2.5 \times 10^{-26}$} & \multicolumn{1}{c}{$1.6 \times 10^{10}$} \\ \hline
\end{tabular}
}
\label{tabla:eficie_def2}
\end{table}

\begin{table}
\tbl{Elipticity ($\epsilon$), GW amplitude ($h_{def}$), GW luminosity ($L_{GW_{def}}$) and efficiency of the mechanism ($\eta _{df})$ for the upper limit of magnetic field ($B_{dip}$) of each system.}
{\begin{tabular}{lccccc}
\hline
\textbf{SYSTEMS} & \begin{tabular}[c]{@{}c@{}}$B_{dip}$ \\  (G)\end{tabular} & $\epsilon$ & $h_{def}$ & \begin{tabular}[c]{@{}c@{}}$L_{GW_{def}}$ \\  (erg/s)\end{tabular} & $\eta _{df}$ \\ \hline
AE Aqr & $5.0 \times 10^{7}$ & $5.1 \times 10^{-9}$ & $6.2 \times 10^{-29}$ & $2.13 \times 10^{21}$ & $1.1 \times 10^{-13}$ \\
AR Sco & $5.0 \times 10^{8}$ & $5.3 \times 10^{-7}$ & $4.6 \times 10^{-28}$ & $1.25 \times 10^{22}$ & $4.02 \times 10^{-12}$ \\
RX J0648 & $1.0 \times 10^{8}$ & $2.8 \times 10^{-10}$ & $9.5 \times 10^{-31}$ & $1.33 \times 10^{20}$ & $1.4 \times 10^{-14}$ \\ \hline
\end{tabular}
}
\label{tabla:eficie_def3}
\end{table}

\subsection{GWs from SGRs/AXPs as fast-spinning WDs}

An alternative model has been proposed for SGRs/AXPs considering they are fast-rotating and magnetized WDs [see e.g. Refs.~\citenum{Malheiro/2012,Coelho/2014,2016IJMPD..2541025L,2016JCAP...05..007M} for further details]. From this perspective, a canonical spin-powered pulsar model can explain the process of energy emission released by dipole radiation in a WD, since they share quite similar aspects \cite{Usov/1988}. In addition, these sources could also be candidates for GW emission, since the high magnetic field can deform the star in a non-symmetrical way, thus generating a variation in the quadrupolar moment of the star. 

Therefore, we consider in this section that SGRs/AXPs are fast-spinning and magnetized WDs which emit GWs due to the deformation caused by their own intense magnetic field. Thus, using Eq. \ref{Ampdef3} and adopting $\kappa\simeq 10$, we calculate the GW amplitude for the 23 confirmed SGRs/AXPs \cite{Olausen/2014} \footnote{For information  about the SGRs/AXPs, we refer the reader to the McGill University's online catalog available at: \url{http://www.physics.mcgill.ca/~pulsar/magnetar/main.html}}, considering these objects as a very massive WD. To do so, we assume three values of mass and their corresponding radius, namely, $M_{WD} = 1.4 M_{\odot }$ ($R_{WD} = 1.0 \times 10^{8}$ cm), 1.2 $M_{\odot }$ ($R_{WD} = 6.0 \times 10^{8}$ cm) and 1.0 $M_{\odot }$ ($R_{WD} = 7.5 \times 10^{8}$ cm) [see Ref. \citenum{Boshkayev2013} for further details about the mass-radius relation]. 

\begin{figure}
\begin{center}
\includegraphics[width=8cm, height=5.3cm]{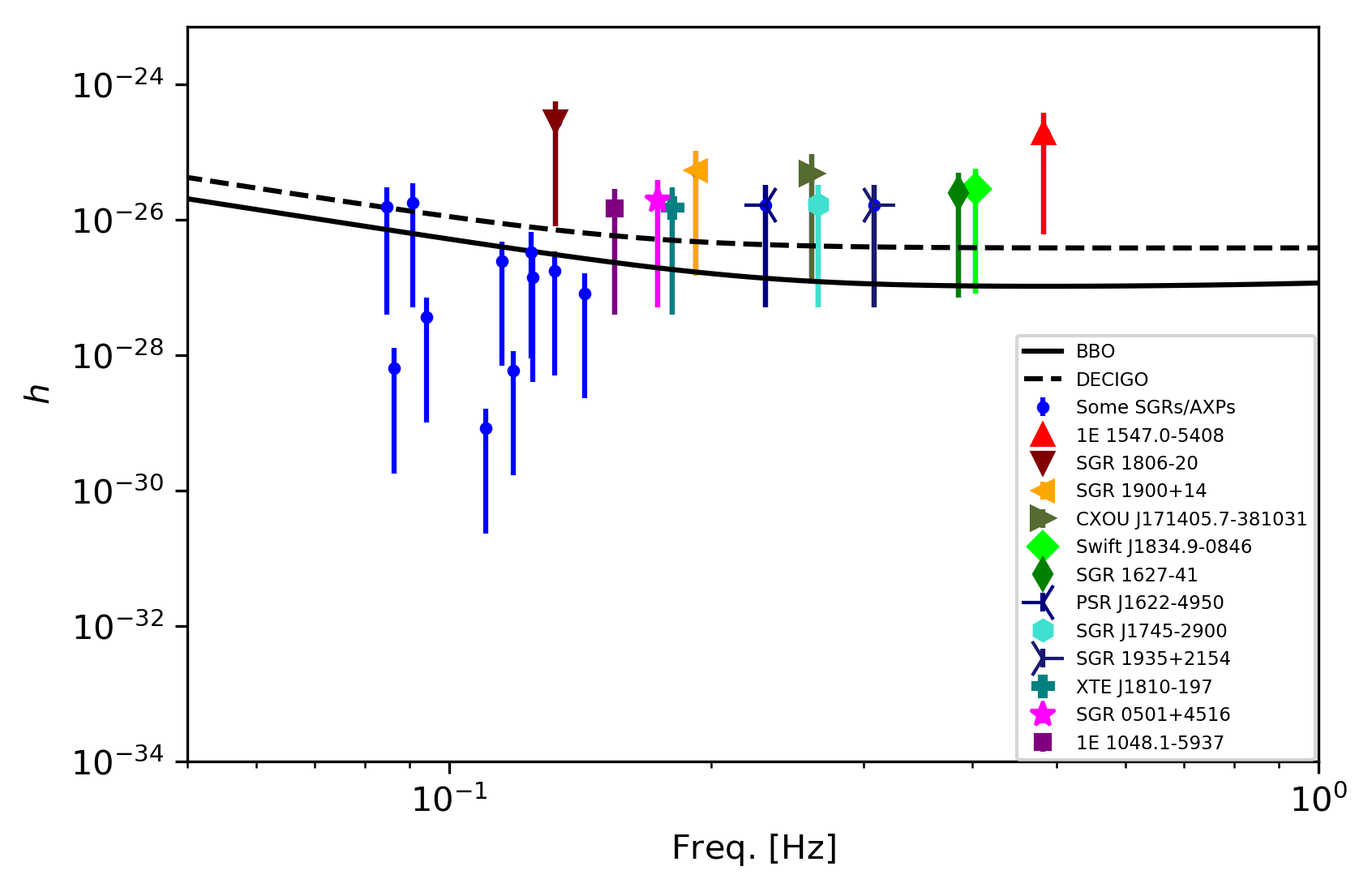}
\caption{GW amplitude as a function of frequency for SGRs/AXPs as fast-spinning and magnetized WDs for masses in the interval  $1.0 M_{\odot} \leq  M_{WD} \leq  1.4 M_{\odot}$, represented by the vertical bars, from top to bottom. The bullets stand for $M_{WD} = 1.2 M_{\odot }$. Also plotted the sensitivity curves for BBO and DECIGO for $SNR = 8$ and integration time $T = 1$ year.}
\label{fig:5} 
\end{center}
\end{figure}

After estimating the amplitude, we placed them on the sensitivity curves of the BBO and DECIGO detectors. Figure \ref{fig:5} shows theses results, such that the GW amplitude is presented as a function of frequency for some SGRs/AXPs. In this Figure, the bullets stands for $M_{WD} = 1.2 M_{\odot }$ and the vertical bars, that crosses the bullets, stands for  $1.0 M_{\odot} \leq   M_{WD} \leq  1.4 M_{\odot} $, from top to bottom. 

\begin{figure}
\begin{center}
\includegraphics[width=8cm, height=5.3cm]{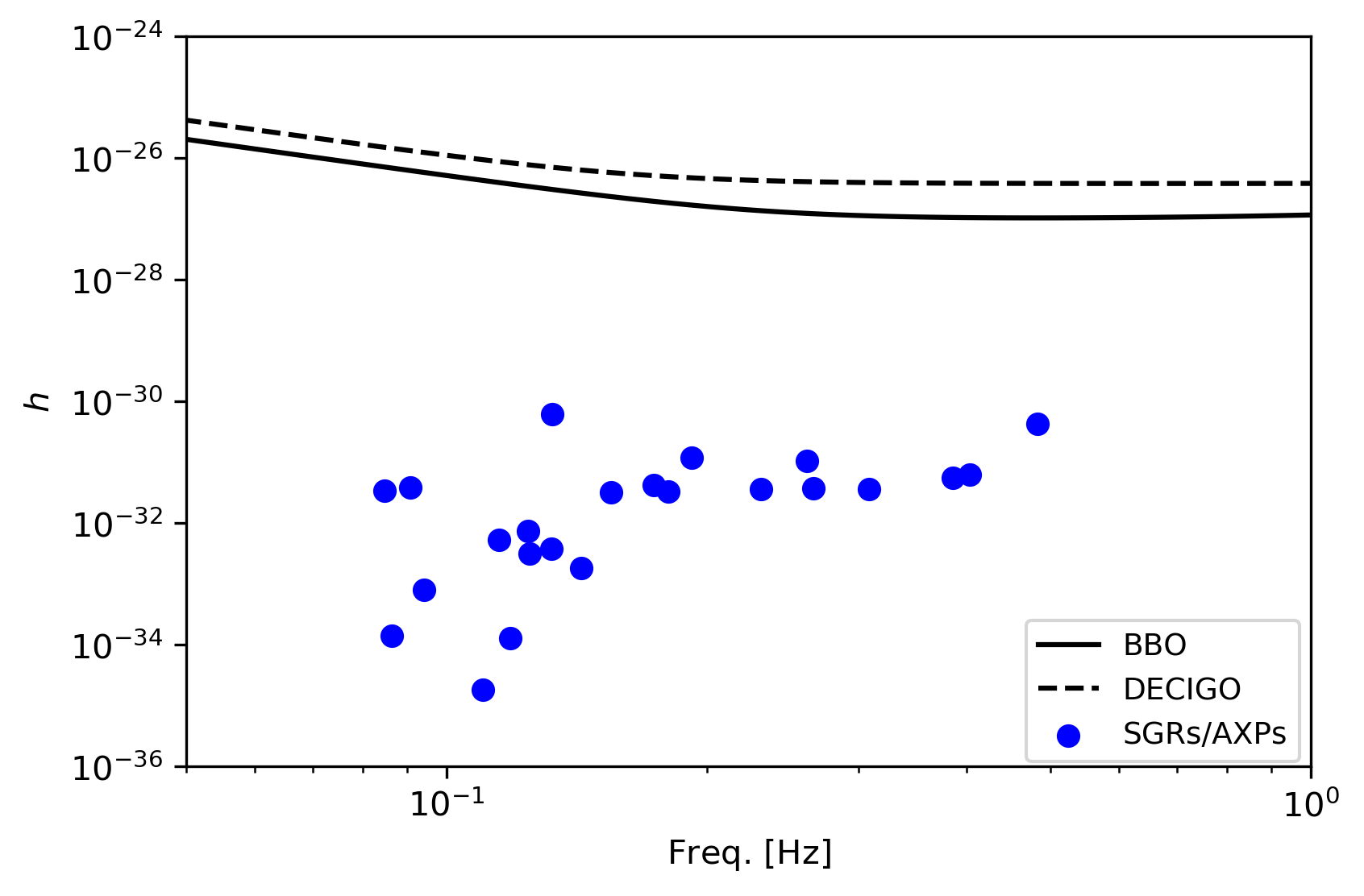}
\caption{GW amplitude as a function of frequency for SGRs/AXPs as NSs. Also plotted the sensitivity curves for BBO and DECIGO for $SNR = 8$ and integration time of $T = 1$ year. We consider a NS of $M = 1.4 M_{\odot}$, radius $R = 10$ km.}
\label{fig:6} 
\end{center}
\end{figure}

Notice that some SGRs/AXPs produce GWs with amplitudes that can be detected by BBO and DECIGO. Some of them, for example 1E 1547.0-5408 and SGR 1806-20, could well be detected for the entire mass range considered, while others would be detected depending on how massive they are.

SGRs/AXPs described as WDs generate GW amplitudes much larger than SGRs/AXPs described in the magnetar model where they are neutron stars [see e.g. Refs. \citenum{duncan1992formation, kaspi2017magnetars} for details about magnetar model]. This is because WDs have moments of inertia four orders of magnitude greater than a neutron star. Consequently, the GW amplitudes generated by these sources as neutron stars are far below the sensitivity curves of BBO and DECIGO [see Fig. \ref{fig:6}]. Therefore, if these space based instruments detect continuous GWs from these sources, this would corroborate the model of fast spinning and magnetic WDs.

\section{Final Remarks}
\label{aba:sec4}

We investigate two mechanisms - accretion of matter and magnetic deformation - for the production of gravitational waves in fast-spinning WDs. These uncommon WDs have high rotation and huge magnetic fields. Also, these stars are considered in an alternative model to describe SGRs and AXPs, where they are characterized as rotation-powered WD pulsars.

Then, we firstly study the following three binary systems: AE Aqr, AR Sco and RX J0648. For the accretion of matter mechanism, our results show that the AE Aqr and RX J0648 are good candidates for BBO and DECIGO if they have an amount of mass accumulated of $\delta m \geq 10^{-5}M_{\odot}$, considering 1 year of integration time and $SNR = 8$. In contrast, AR Sco is unlikely to be detected because it is needed a large amount of mass accumulated in the magnetic pole of this WD. 

Now, regarding the magnetic deformation mechanism, to emit gravitational radiation with amplitudes that are detectable by BBO, for example, the three binary systems studied require that the WD has a magnetic field above $\sim 10^{9}$ G. Nevertheless, these WDs are inferred to have magnetic fields with intensity around two orders of magnitude smaller. Moreover, we also conclude that gravitational radiation has an irrelevant contribution to the spindown luminosity of these systems for both mechanism.

Still, taking into account the magnetic deformation mechanism, we investigate the SGRs/AXPs as rotation-powered WD pulsars assigning a mass range $1.0 M_{\odot} \leq   M_{WD} \leq  1.4 M_{\odot}$ for these objects. We conclude that a possible detection of continuous GWs coming from SGRs/AXPs would be a good indication that could corroborate the WD pulsar model, as for the neutron stars description, they are far below the BBO and DECIGO sensitivity curves.

\section*{Acknowledgements}
M.F.S. thanks CAPES for the financial support. J.C.N.A. thanks FAPESP (2013/26258-4) and CNPq (308367/2019-7) for partial financial support. J.G.C. is likewise grateful to the support of  CNPq (421265/2018-3 and 305369/2018-0), FAPESP Project No. 2015/15897-1, and NAPI (Fenômenos Extremos do Universo) of Fundação de Apoio à Ciência, Tecnologia e Inovação do Paraná.

\bibliographystyle{ws-procs961x669}
\bibliography{ws-pro-sample}

\end{document}